\documentstyle[aps,twocolumn,epsfig,rotating]{revtex}
\begin{document}

\draft \preprint{}

\begin{title} {\bf Microscopic Phase Separation in the Overdoped
Region of High-$T_{c}$ Cuprate Superconductors\/}
\end{title}

\author{Y.J.~Uemura}

\address{Department of Physics, Columbia University, 
New York, NY 10027, USA}

\vspace{-5truemm}

\date{\today}
\maketitle

\begin{abstract}

\noindent 
We propose a phenomenological model for 
high-$T_{c}$ superconductors (HTSC) assuming: (1) a microscopic phase
separation between superconducting and normal-metal areas in 
the overdoped region; and 
(2) existence of a homogeneous superconducting phase 
only below the pseudo-gap $T^{*}$ line, which shows a sharp reduction 
towards $T^{*}\sim 0$ at a mildly overdoped critical concentration $x_{c}$.
This model explains anomalous doping
and temperature dependences of $n_{s}/m^{*}$ 
(superconducting carrier density / effective mass) observed in 
several overdoped HTSC systems.
We point out an analogy to superfluid $^{4}$He/$^{3}$He films, 
and discuss an energetic origin of microscopic phase separation. 
\end{abstract} 


\narrowtext

\noindent
Since the discovery of high-$T_{c}$ cuprate superconductors (HTSC),
accumulated studies have revealed unusual phenomena 
in the underdoped (UD)
region, such as, correlations between $T_{c}$ and $n_{s}/m^{*}$
(superconducting carrier density / effective mass) at $T\rightarrow 0$ 
shown by 
measurements of the magnetic field penetration depth 
$\lambda$ [1], and the pseudo-gap phenomena [2].  
These results stimulated development of 
various theories / models for condensation, including
Bose-Einstein (BE) to 
BCS crossover [3,4], phase fluctuations [5,6],
XY-model [7] and BE condensation [8].  
These models [3-8] assume the existence of pre-formed charge 
pairs above $T_{c}$.  
Different pictures, however, assume single charge above $T_{c}$
based on the resonating-valence-bond concepts [9,10].  

Several anomalous results have been found also in the overdoped (OD) region: 
$\mu$SR studies on Tl$_{2}$Ba$_{2}$CuO$_{6+\delta}$ (Tl2201) [11,12] revealed 
that $n_{s}/m^{*}$ at
$T\rightarrow 0$ decreases with increasing hole doping.  
This tendency has been observed 
subsequently in thin film La$_{2-x}$Sr$_{x}$CuO$_{4}$ (LSCO) [13] and bulk 
YBa$_{2}$Cu$_{3}$O$_{y}$
(YBCO) [14] systems in the OD region.
The ``coherence peak'' in ARPES spectra [15] also follows 
this behavior of
$n_{s}/m^{*}$ in the OD region.
Meanwhile, Tallon and Loram [16] noticed a sharp reduction of 
the pseudo-gap $T^{*}$ line in the temperature $T$ vs. hole-concentration 
$x$ phase diagram
heading towards $T^{*} \sim 0$ at a critical concentration
$x_{c} \sim 0.19$ holes/Cu which lies in the mildly OD region.  
In view of the existence of superconductivity
in $x > x_{c}$ where the $T^{*}$ line does not exist, 
Tallon and Loram [16] have
advocated a view point that the pseudogap phenomena is not representing
pre-cursor superconducting phenomena, but should be ascribed to a
tendency towards a ground state competing with superconductivity.
  
We have, however, suggested another possible view point
in which $T^{*}$ is ascribed to a signature of a gradual pre-formation of 
pairs [4] while the anomaly in the OD region to a phase separation 
[4,11,17,18].
In the present paper, resorting to a model-calculation of $n_{s}/m^{*}$,
comparison with experimental results, a crude estimate of competing 
electrostatic and condensation/pairing energies, and analogy to
$^{4}$He/$^{3}$He films, we demonstrate that
our picture with microscopic phase separation between superconducting
and normal-metal regions can 
quantitatively account for several anomalous results in the OD region.
We have pointed out an 
analogy to He films in a recent conference [18].
The present model introduces a new type of possible charge 
heterogeneity to 
HTSC systems, in addition to other known examples, such as 
charge/spin stripes [19].  

Figure 1(a) shows the results of $T_{c}$ versus the muon spin 
relaxation rate $\sigma \propto n_{s}/m^{*}$ at $T\rightarrow 0$ 
of YBCO [1,14], Zn-doped YBCO ($y=6.7$) [20] and overdoped Tl2201 [11] systems.
$T_{c}$ increases with increasing hole doping following a linear line
in the UD, a saturation in the optimal $T_{c}$ (OPT), and a recurring behavior
in the OD region.  This figure for HTSC systems exhibits a striking 
resemblance to 
the corresponding plot for superfluid $^{4}$He and $^{4}$He/$^{3}$He 
films in non-porous [21] and porous media
[22], shown in Fig. 1(b).
Here, the superfluid transition temperature $T_{c}$ is plotted versus
the superfluid density  
$n_{s2d}/m^{*} \equiv 4n_{b2d}/m_{b}$ at
$T\rightarrow 0$,
where $n_{b2d}$ and $m_{b}$ represent the 2-dimensional area
density and mass of superfluid He atoms (bosons) and $n_{s2d}=2n_{b2d}$ and 
$m^{*} = m_{b}/2$ represent corresponding values in fermion terminology.

Simple hole doping in underdoped HTSC can be viewed as analogous to 
He films on Mylar [21]. 
Zn-doped YBCO [20] can be compared to He films in porous Vycor glass [22],
since a non-superconducting / non-superfluid area is formed around Zn / pore
surface as a ``healing region'', while $T_{c}$ is determined by the 
remaining superfluid density in both cases.  We assumed a normal 
region with the area of $\pi\xi_{ab}^{2}$ ($\xi_{ab}$ is the in-plane
coherence length) on the CuO$_{2}$ plane around each Zn, and showed
that the reduction of $n_{s}/m^{*}(T\rightarrow 0)$ with increasing Zn
concentration in YBCO and LSCO can 
be explained by this ``swiss cheese model'' [20].  This hypothesis has been
confirmed by scanning
tunneling microscopy (STM) measurements [23].       
Such a coexistence of superfluid / normal regions can be viewed as 
an example of a ``microscopic phase separation''.  We also found a similar
situation in HTSC superconductors with static stripe spin freezing [17].

When $^{3}$He is mixed into $^{4}$He, the superfluid transition temperature
is reduced with increasing $^{3}$He fraction $p_{3}$, as shown by the phase
diagram in the inset of Fig. 1(b).  There is a large region of phase
separation between boson-($^{4}$He)-rich and fermion-($^{3}$He)-rich 
liquids.  In a bulk mixture, the lighter fermion-rich liquid in the 
upper part of a container does not mix with the boson-rich superfluid.  
Adsorption of $^{4}$He/$^{3}$He mixture onto fine alumina
powder [24] presumably keeps boson-rich
and fermion-rich liquids coexisting in a microscopic length scale,
resulting in a superfluid film in the full range of $p_{3}$.
 
The results of $^{4}$He/$^{3}$He mixture on alumina powder [24] 
in Fig. 1(b)
exhibit a roughly-linear relation between $T_{c}$ and the 
superfluid density.  This behavior  
is analogous to that of the overdoped HTSC systems in Fig. 1(a).  
Both of these cases represent the response of 
superfluidity / superconductivity to increasing fermions ($^{3}$He and holes).
In all the cases of He films in Fig. 1(b), $T_{c}$ is determined by the 
area-averaged superfluid density.  Similarity between Fig. 1(a) and 1(b)
suggests the possibility that $T_{c}$ in HTSC systems may also be determined
by $n_{s}/m^{*}$ at $T\rightarrow 0$ 
averaged over a length scale of several times $\xi_{ab}$.

Figure 2(a) compares the doping dependence of $n_{s}/m^{*}$ from $\mu$SR [11] 
with that of the ``gapped'' response $\gamma_{s}$ in the T-linear term 
$\gamma$ of the specific heat [25], observed in Tl2201.
The normal state value $\gamma_{n}$ in Tl2201 above $T_{c}$
is virtually independent of doping, which implies  
no doping dependence of $m^{*} \propto \gamma$.
By the broken line, we show a projected variation 
of the normal-state carrier density / mass, 
$n_{n}/m^{*} \propto x/\gamma_{n} \propto x$.  
Departure of $n_{s}/m^{*}$
from $n_{n}/m^{*}$ suggests that only a part of normal-state carriers
form superfluid.   We cannot ascribe this departure to 
the scattering effect, since
the transport mean-free path {\it l\/} of Tl2201 is much longer
than $\xi_{ab}$, even for a 
highly overdoped sample with $T_{c} \sim 20$ K [26].  
The BCS theory with retarded interaction
cannot account for this phenomenon [4,11].
The common behavior of $n_{s}/m^{*}$ and $\gamma_{s}$
in Fig. 2(a) suggests a possibility that the departure of 
$n_{s}$ from $n_{n}$ maybe related to a volume effect. 

Motivated by these observations, we propose a phase diagram of HTSC
systems shown in Fig. 3, where the $T^{*}$ line is ascribed to
pair formation, and the OD region is characterized by a 
phase separation
between the hole-poor superconducting region with $x_{s}(T)$ along
the $T^{*}$ line and the non-superconducting hole-rich 
region with $x_{f}(T)$ along the $T_{c}$ line in the OD region.
We assume a microscopic phase separation via formation of 
non-superconducting regions with the length scale of $\xi_{ab}$, analogous
to the ``swiss cheese model'' in Zn-doping, as illustrated in 
Fig. 3.  For simplicity, we perform a model calculation assuming
(1) parabolic shape of the $T_{c}$ line which has maximum at 
$T_{c}(x_{opt}=0.15)\equiv T_{c}^{max}$ and intersects with 
$x$ axis at $x_{max} = 0.27$ and $x_{min} = 0.03$; and (2) linear 
$T^{*}$ line connecting $T^{*}(x_{opt}) = T_{c}^{max}$ and 
$T^{*}(x_{c} = 0.19) = 0$.
For a given hole concentration $x_{1}\geq x_{opt}$, 
cooled down from high temperature,
we assume the system to separate at $T\leq T_{c}$ into the superconducting 
liquid with $x_{s}(T)$ and normal liquid with
$x_{f}(T)$, having the volume fraction of $p_{s}$ and $p_{f}=1-p_{s}$, 
respectively, where $x_{1} = (x_{s}p_{s}+x_{f}p_{f})$.  Below
$T = T^{*}(x_{1})$, the total volume becomes superfluid.   
  
To consider an energetic origin of phase separation, let us
imagine a capacitor 
having an area $A\equiv\pi\xi_{ab}^{2}$ for $\xi_{ab} = 15$ \AA\ and
thickness comparable to the average interlayer distance 
$c_{int} \sim 6$ \AA, charged with $+/-$\ $Q \sim 2e$ given by the 
deviation from average charge $(x_{max}-x_{c})/2 = 0.04$
[holes/Cu] multiplied to the number of Cu atoms (48) in the area $A$.
An assembly of alternating charge layers stacked along the c-axis 
direction can be 
expressed by sets of such capacitors with the charge $+/-$\ $Q/2 \sim e$ on 
each plate.  
The electrostatic energy to have one such capacitor is  
$E = (Q/2)^{2}/C \sim 0.8/\epsilon \sim 0.1$ eV,
where $\epsilon \sim 10$ represents an effective static dielectric constant 
due to atoms and ions between the CuO$_{2}$ planes,
and $C$ denotes the capacitance.
Imagine hole-poor and hole-rich regions, adjacent to each other, each 
having area $A$ on a given CuO$_{2}$ plane.
To create this situation we need energy $2E$ per area $2A$ on a CuO$_{2}$ plane.
In view of further energy saving via Madelung potential,
we estimate the actual electrostatic energy $E_{Coulomb}$ to be roughly
$\sim 0.1$ eV.

This energy cost for 
charge disproportionation competes with the gain of 
condensation and pairing energies
$E_{CP}$ for having the hole-poor
area $A$ with $q_{s} = A \times x_{c} \sim 9e$ charges paired and condensed.
Assuming $\Delta =1.7k_{B}T_{c}$ energy gain per charge and $T_{c} \sim 90$K, 
we obtain $E_{CP} \sim 0.12$ eV. 
For BE condensation, $\Delta$ should be replaced by the sum of the 
pairing energy
$\propto k_{B}T^{*}$ and the condensation energy $\propto k_{B}T_{c}$, 
while for BCS condensation $E_{CP}$ has to be 
multiplied by the ratio 0.1-0.2 of 
$\Delta$ to the effective Fermi energy $\epsilon_{F} \sim 0.2$ eV.
Even in the purely superconducting region at $x \leq x_{c}$, the system might  
spontaneously introduce some charge heterogeneity within 
$x_{min} < x <x_{c}$ to gain the pairing energy.

In LSCO, for example, the random spatial distribution of Sr$^{2+}$ 
will further    
reduce $E_{Coulomb}$ substantially.  The area $A$ with 48 Cu atoms
is associated with $N_{Sr} \sim$ 8-10 Sr ions in the OPT region.
We expect $\sqrt{N_{Sr}} \sim 3$ random fluctuations in this number,
which would promote natural formation of hole-rich and hole-poor regions.  
The combination of these effects can make $E_{CP} > E_{Coulomb}$, 
and possibly result in a microscopic phase separation to minimize 
the total energy.
In the capacitor argument, both $E_{CP}$ and $E$ are proportional to $A$.
This feature does not give any preference for the magnitude of $A$.  
The lower limit of $A$ is related to $\xi_{ab}$ and the discreteness of 
the charge.  The upper limit of $A$ may be related to the energy
gain via $\sqrt{N_{Sr}} \propto \sqrt{A}$ and loss of percolation /
proximity effect of 
superconducting regions for larger $A$. 

Using the model shown in Fig. 3, we calculated the superfluid 
density $n_{s}(T=0)$ as $n_{s}=x$ for $x\leq x_{c}$ and
$n_{s}=x_{c}p_{s}$ for $x_{c}<x<x_{max}$, where $p_{s}$ and $p_{f}$ are
volume fractions of hole-poor liquid with $x_{c}$ and hole-rich liquid
with $x_{max}$, $p_{s} = 1-p_{f}$, and $x = x_{c}p_{s} + x_{max}p_{f}$.  
We assumed $T_{c}^{max}=90$ K, $x_{c}=0.19$ and $x_{max}=0.27$, and show the 
results in the inset of Fig. 1(a).   
In Fig. 2(b), we also compare the 
published results for YBCO [14] and the variation of $n_{s}$ from
our calculation.  The good agreements of calculation and 
experiments demonstrate that the phase
separation can account for the ``recurring'' behavior
of $n_{s}/m^{*}(T\rightarrow 0)$
in the OD region and its anomalously sharp change 
around $x_{c}$.

The $\mu$SR results of $n_{s}/m^{*}$
in Tl2201 [11] and overdoped LSCO systems [27] exhibit anomalous 
temperature dependence characterized by increasing sharpness near $T \sim 0$ 
with increasing doping,
as shown in Fig. 4(a) and 4(c).  
Unfortunately, the $\mu$SR results were 
obtained on ceramic specimens, which often show deviation from
predicted variation for d-wave energy gap even in the OPT region. 
This feature prohibits detailed comparison with theories.  
However, we performed model calculation in the following assumptions/steps:
(1) the thermal pair-breaking effect within the hole-poor superconducting
region can be represented by the experimental results 
$\sigma_{OPT}(T)$ obtained 
for specimens in the (nearly) OPT region with highest $T_{c}$;
(2) $n_{s}/m^{*}(T)$ of OD specimens can be calculated by multiplying
the hole concentration of the hole-poor superconducting region
$x_{s}(T)$ at $T$ with its volume fraction $p_{s}(T)$, and 
further by $\sigma_{OPT}(T)/\sigma_{OPT}(T=0)$.  The results of this
calculation,
shown in Fig. 4(b) for Tl2201 and Fig. 4(c) for LSCO,
reproduce
the observed temperature/doping dependence very well.  

In overdoped cuprates, we are not sure whether the non-superconducting
hole-rich regions are spatially pinned to charge randomness, or
they are dynamically fluctuating.  STM 
measurements would be most effective to study this feature.
A similar departure of $n_{s}/m^{*}$ from $n_{n}/m^{*}$
was also found in the 2-d
organic superconductor (BEDT-TTF)$_{2}$Cu(NCS)$_{2}$ [28]. 
It will be interesting to  
investigate the applicability of our model to BEDT and
other superconducting systems.
In conclusion, we have presented a model with microscopic phase separation 
to account for the anomalous behavior observed in overdoped HTSC systems.
The present picture provides a possible
way to reconcile the existence of superconductivity in the OD
region with the sharp reduction of the $T^{*}$ line.  

This work was supported by NSF-DMR-98-02000 and 01-02752,
and US-Israel Binational Science Foundation.  
The author acknowledges useful discussions
with M. Randeria and O. Tchernyshyov.

\clearpage

\begin{figure}[!T]
\begin{center}
\mbox{\epsfig{file=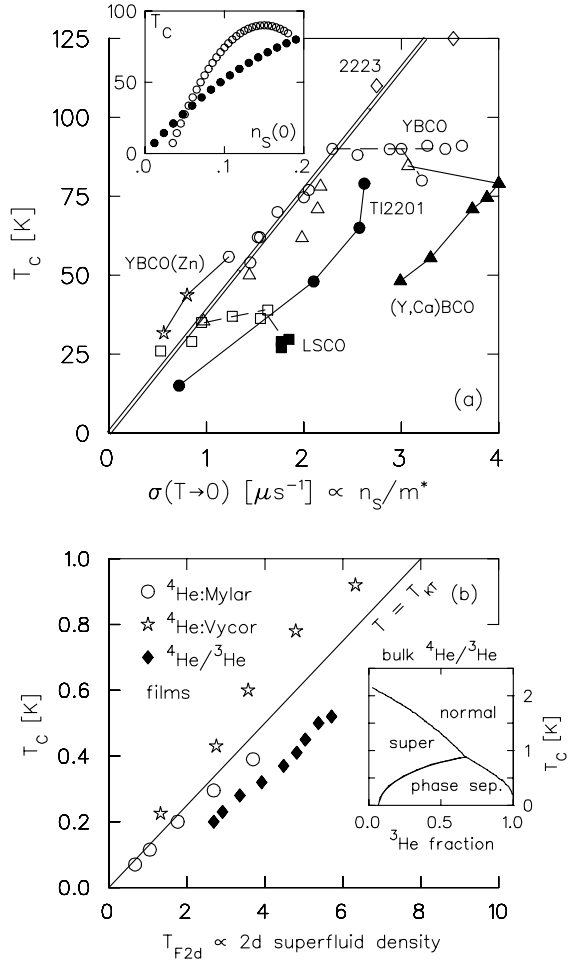,height=5.00in}}
\vskip.5pc
\caption{(a) Plot of $T_{c}$ versus muon spin relaxation rate
$\sigma(T\rightarrow 0) \propto n_{s}/m^{*}$
of HTSC systems in the UD-OPT (open symbols) and OD (closed
symbols) regions [1,11,14,20]. Inset shows $T_{c}$ versus $n_{s} = x_{c}p_{s}$ 
from the present model for $T_{c}^{max}$ = 90 and $x_{c}$ = 0.19 [holes/Cu].
(b) Plot of $T_{c}$ versus the 2-d Fermi temperature
$T_{F}$ ($\propto$ 2-d superfluid density) for $^{4}$He and 
$^{4}$He/$^{3}$He mixture films adsorbed on Mylar, Vycor and alumina
powders [21,22,24]. The solid line represents Kosterlitz-Thouless
transition temperature $T_{KT}$.}
\label{fig1ab}
\end{center}
\end{figure}

\begin{figure}[T]
\begin{center}
\rotatebox{0}{\mbox{\epsfig{file=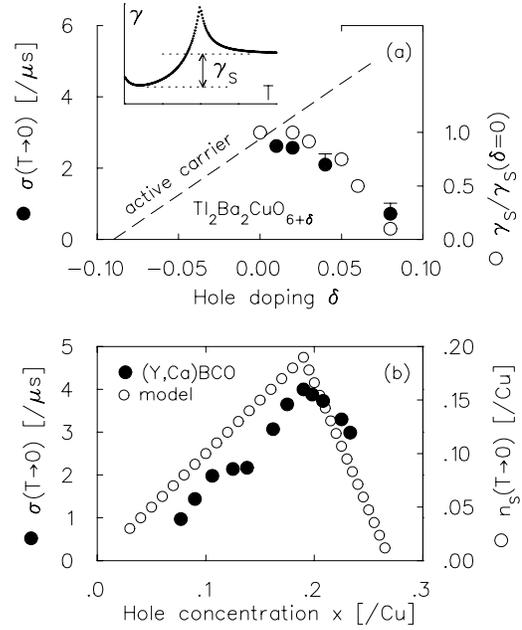,height=3.250in}}}
\vskip.5pc
\caption{(a) Muon spin relaxation rate $\sigma(T\rightarrow 0)\propto 
n_{s}/m^{*}$
(closed circles) [11] and the ``gapped'' response $\gamma_{s}$ 
in the linear-term of the specific heat (open circles) [25]
in Tl2201.  The broken
line illustrates a projected variation of $n_{n}/m^{*}$.
(b) $\sigma(T\rightarrow 0)\propto 
n_{s}/m^{*}$ in YBCO [14] (closed circles) and $n_{s}(T\rightarrow 0)$
= $x_{c}p_{s}$ for $x\geq x_{c}$ and $x$ for $x<x_{c}$ (open circles)
from our model plotted versus hole concentration $x$ [/Cu].}
\label{fig2ab}
\end{center}
\end{figure}

\begin{figure}[T]
\begin{center}
\rotatebox{0}{\mbox{\epsfig{file=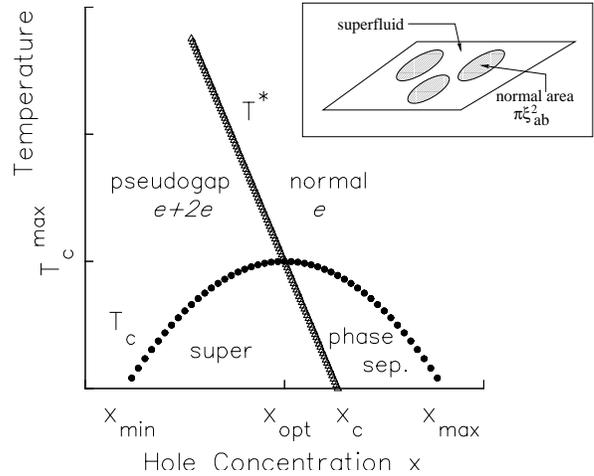,height=2.5in}}}
\vskip.5pc
\caption{Proposed phase diagram of HTSC systems.  For model 
calculation, the $T_{c}$ curve is approximated by a parabola and
the $T^{*}$ by a line. The inset illustrates proposed microscopic
phase separation in the OD region.}
\label{fig3}
\end{center}
\end{figure}

\clearpage

\widetext
\begin{figure}[T]
\begin{center}
\rotatebox{270}{\mbox{\epsfig{file=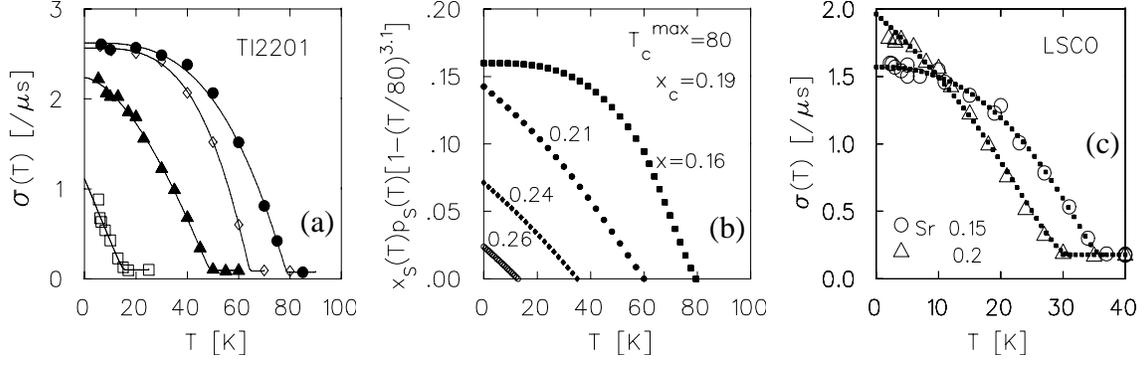,height=6.5in}}}
\vskip.5pc
\caption{
(a) Temperature dependence of the muon spin relaxation rate $\sigma$ 
observed in overdoped Tl2201 [11].
(b) Model calculation of $n_{s}(T) 
\equiv x_{s}(T)p_{s}(T)(1-(T/T_{c})^{\beta})$
with $\beta = 3.1$
for $T_{c}^{max}$ = 80 K and $x_{c}$ = 0.19.
(c) Experimental results of $\sigma(T)$ in LSCO [27],
compared with the model calculation for $T_{c}^{max}$ = 36.3 K,
$x_{c}$ = 0.193, and $n_{s}$ given as in (b).  $\beta$ = 2.37
was obtained by fitting the observed results for Sr 0.15,
while the curve for Sr 0.2 represents our model for $x$ = 0.195.}
\label{fig4abc}
\end{center}
\end{figure}

\end{document}